\documentclass{emulateapj}
\usepackage{natbib}

\shorttitle{The Dark Halo in NGC~821}
\shortauthors{Forestell and Gebhardt}

\begin{document}

\title{Hobby-Eberly Telescope Observations of the Dark Halo in
NGC~821\altaffilmark{1}}

\author{Amy D. Forestell and Karl Gebhardt} \affil{Department of
Astronomy, University of Texas at Austin, C1400, 1 University Station,
Austin, TX 78712} \email{amydove@astro.as.utexas.edu,
gebhardt@astro.as.utexas.edu}

\altaffiltext{1}{Based on observations obtained with the Hobby-Eberly
Telescope, which is a joint project of the University of Texas at
Austin, the Pennsylvania State University, Stanford University,
Ludwig-Maximilians-Universit\"at M\"unchen, and
Georg-August-Universit\"at G\"ottingen.}

\begin{abstract}

We present line-of-sight stellar velocity distributions of elliptical
galaxy NGC~821 obtained to approximately $100\arcsec$ (over 2
effective radii) with long-slit spectroscopy from the Hobby-Eberly
Telescope. Our measured stellar line-of-sight
velocity distributions are larger than the planetary nebulae
measurements at similar radii.  We fit axisymmetric orbit-superposition models with a range
of dark halo density profiles, including two-dimensional kinematics at
smaller radii from SAURON data.  Within our assumptions, the
best-fitted model gives a total enclosed mass of $2.0\times 10^{11}
M_{\sun}$ within $100\arcsec$, with an accuracy of 2\%; this mass is
equally divided between halo and stars.  At $1R_e$ the best-fitted
dark matter halo accounts for $13\%$ of the total mass in the galaxy.
This dark halo is inconsistent with previous claims of little to no
dark matter halo in this galaxy from planetary nebula measurements.
We find that a power-law dark halo with a slope 0.1 is the best-fitted
model; both the no dark halo and NFW models are worse fits at a
greater than 99\% confidence level.  NGC~821 does not appear to have
the expected dark halo density profile.  The internal moments of the stellar
velocity distribution show that the model with no dark halo is
radially anisotropic at small radii and tangentially isotropic at
large radii, while the best-fitted halo models are slightly radially
anisotropic at all radii.  We test the potential effects of model
smoothing and find that there are no effects on our results within the
errors.  Finally, we run models using the planetary nebula kinematics
and assuming our best-fitted halos and find that the planetary nebulae
require radial orbits throughout the galaxy.
\end{abstract}


\section{Introduction}

Cold dark matter is now accepted as an integral part of our universe,
and recent observations have continued to provide support for its
existence \citep{kom08}.  Part of the picture of the universe is that
galaxies are surrounded by massive dark matter halos in which they
formed \citep{whi78,blu84}.  Recently cosmological simulations have
become detailed enough to reach the level of individual galaxy
formation \citep{naa07,gov07}, and comparisons with data can help
further constrain cosmological theory \citep{ost03}.  Indeed, spiral
galaxy rotation curves are one of the strongest pieces of
observational evidence for the existence of dark matter
\citep{van85,per96,sof01}.  It is also important to study the dark
halo structure of elliptical galaxies because of their different
formation and evolution.  However it is more difficult to measure dark
matter in elliptical galaxies because of a lack of tracers at large
radii where dark matter is thought to dominate.  The best way to
measure the underlying gravitational potential is to use kinematics
from the stellar population, but this has been limited due to the
faintness of stellar light in the outer regions of galaxies
\citep{ger01}.  Dark matter in elliptical galaxies has therefore been
studied in other ways, such as via X-ray emission \citep{loe99,mat03},
gravitational lensing \citep{kee01,man08}, and using individual stellar
velocities as in nearby dwarf spheroidals \citep{mat98,kle02}.  In
order to study a more representative sample of galaxies, tracers such
as globular clusters \citep{zep00,pie06} and planetary nebulae
\citep{men01,rom03,coc09} are used used to probe the outer parts of
elliptical galaxies, though it is difficult to get a significant
sample size.  Additional issues arise with these tracers, as discussed
below, such as understanding their radial profile.  With larger
telescopes we are now able to measure stellar kinematics from
integrated light to larger radii, thus closing the gap between stars
and the large-radii tracers.

Meanwhile dynamical models of galaxies have also improved.  Rather
than previous spherical models that use analytic distribution
functions \citep[DFs;][]{ger01}, orbit-based axisymmetric models are
now available.  These fully general models, based on the technique of
\citet{sch79}, provide detailed information on the orbital structure
of the galaxy, including the DF and its projections, such as velocity
anisotropy.  Orbit-based models are now frequently applied to galaxies
for studies of both dark halos and central black holes
\citep{rix97,van98,cre99,geb00,cap02,ver02b,geb03,tho05,geb09}.

The elliptical galaxy NGC~821 is an example in which the use of
large-radii tracers has provided an intriguing result.  \citet{rom03}
study the dark halo of NGC~821 using approximately 100 planetary
nebula velocities and find small line-of-sight velocity dispersions
that are consistent with little or no dark halo.  \citet{dek05} use
disk galaxy merger simulations to show that large anisotropies can be
created in the resulting elliptical galaxies, and that this anisotropy
in combination with the different density profile of a young
population could explain how the low dispersions from planetary
nebulae measurements are also consistent with typical dark matter
halos.  Our study uses deep long-slit spectroscopy of NGC~821 from the
9.2-meter Hobby-Eberly Telescope to obtain stellar kinematics to
greater than 2 effective radii in hopes of further constraining the
dark halo of this galaxy.

\citet{wei09} model NGC~821 using data from SAURON, both at
small radii (which we include in our analysis) and newer data at large
radii. We find similar results both for the kinematics and for the
dark halo properties. Comparison between the two studies is presented
in their paper and within this paper.

NGC~821 is classified as an E6? \citep{dev91}.  It has disky isophotes
\citep{lau85,ben88} and a power-law central surface brightness profile
\citep{rav01}.  The blue absolute magnitude is $-20.27$ \citep{tra00}.
We use a distance of 23.44 Mpc taken from \citet{cap06}, which adjusts
the \citet{ton01} values for the new Cepheids zero-point of
\citet{free01}.  NGC~821 is not detected in H$\alpha$ \citep{mac96} or
OIII \citep{sar06}.  Point source and diffuse X-ray emission has been
detected but there is no evidence for hot gas \citep{pel07a, pel07b}.
NGC~821 is considered a fast rotator \citep{cap07,ems07}.
\citet{pro05} find that NGC~821 has very strong age and metallicity
gradients, from $\sim$4 Gyr and 3 times solar in the center to
$\sim$12 Gyr and less than \onethird\ solar at $1R_e$.  They conclude that NGC~821
has experienced a recent ($\sim$1-4 Gyr ago) burst of star formation,
most likely from in-situ gas and perhaps triggered by the accretion of
a small satellite galaxy.  This may be an indication that there are
young planetary nebulae in this galaxy.

\S \ref{obs} describes the observations and data reduction; in \S
\ref{kin} we describe the kinematic extraction; the dynamical models
are described in \S \ref{mod}; we present our results in \S \ref{res}
and give conclusions in \S \ref{disc}.

\section{Observations and Data Reduction}\label{obs}

Long-slit spectra were taken with the Low-Resolution Spectrograph
\citep{hil98} on the Hobby-Eberly Telescope.  We use the g2 grism and
1\arcsec~by 4\arcmin~slit over the wavelength range 4300-7300\AA.
This setup gives a resolving power of 1300 or a full-width
half-maximum (FWHM) resolution of about 230 km s$^{-1}$.  Measurements
of night sky line widths show that we can measure dispersions to about
110 km s$^{-1}$.  The CCD frame (binned $2\times2$) has a plate scale
of 0.47\arcsec/pix spatially and 2\AA/pix spectrally.  The gain is
1.832 $e^-$ ADU$^{-1}$ and readout noise is 5.10 $e^-$.  We use the
Schott Glass blocking filter GG385, which has a half-power point of
the transmission around 385 nm.

NGC~821 was observed over eight nights in November 2003 for a total
exposure time of approximately 5.5 and 2.3 hours on the major and
minor axes respectively.  Cadmium and Neon calibration lamp exposures
and white light illumination flat fields were taken each night.

The data reduction uses standard techniques.  First we overscan
correct and trim the images.  Then we apply a flat correction using a
normalized flat frame, taken from averaged instrumental flats obtained
each night of observations.  Next we rectify the images along the
spatial axis using the calibration lamp lines as a reference.

For sky subtraction, we use the region of the slit that is furthest
from the galaxy center. Since we only have a 4\arcmin\ slit, there
will be some galaxy light in the region where we select sky. However,
the surface brightness profile extends out 350\arcsec\ so we can
accurately calculate the amount of galaxy in our background
region. For our last extracted spectrum (at 90\arcsec), the amount of
galaxy light that we are including as background light is about 15\%
of the galaxy light for that last extraction. We have run simulations
in order to determine whether this amount of contamination has an
effect on the extracted kinematics. We take a high signal-to-noise
galaxy spectrum and subtract off 15\% of itself, and then extract the
kinematics. Only for very high S/N does this amount have an effect and
for the S/N for this dataset (as described below), we find no
significant effect.

\section{Kinematics}\label{kin}

We extract the spectra in radial bins along the major and minor axes.
Because the seeing is approximately 2$\arcsec$ we set the central bins
to 5 pixels (2.35$\arcsec$).  The outer bins are sized to obtain
sufficient signal for kinematic analysis.  Along the minor axis the
spectra from either side of the galaxy were averaged at each radius before kinematic analysis.
Along the major axis, the center of the galaxy was near the edge of
the chip so only one side was extracted.  Our farthest radial bin
extends to 99$\arcsec$ on the majoraxis and 45$\arcsec$ on the minor axis, corresponding to a V-band surface brightness of 21.8 and 23.5 respectively.  There is a faint object centered at 115$\arcsec$ along the major axis that prevents a further radial bin from being used.  The radial extent in effective radii depends on the value
of $R_e$ used.  The measured $R_e$ of NGC~821 varies throughout the
literature: 50$\arcsec$ \citetext{RC3}, 45$\arcsec$ \citep{fab89},
39$\arcsec$ \citep{cap06}, 36$\arcsec$ \citep{tra00}, and
16.7-18.3$\arcsec$ \citep{ben88}.  For the purpose of this discussion
we adopt an $R_e$ of 45$\arcsec$.  Thus our data extend to
approximately $1 R_e$ along the minor axis and $2 R_e$ along the major
axis.

We do not flux calibrate the spectra, and thus we remove the continuum
in each spectrum.  We fit the local continuum by finding the biweight
\citep{bee90} in windows as described in \citet{pin03}.  The
wavelength solution comes from the Cd and Ne calibration lamps.

We obtain a nonparametric line-of-sight velocity distribution (LOSVD)
by deconvolving the galaxy spectrum with a set of stellar template
spectra using the maximum penalized likelihood technique of
\citet{geb00}.  Tests of this technique are given in \citet{pin03}.
There are 30 evenly-spaced velocity bins of 54 km s$^{-1}$ that
represent the LOSVD.  We vary the height in each bin and the weights
of each template star to find the best match to the galaxy spectrum from each radial range.
We use nine stellar templates with types ranging from G dwarf to M
giant from \citet{lei96}, convolved to our spectral resolution.

\begin{figure}
\includegraphics[angle=270,scale=.3]{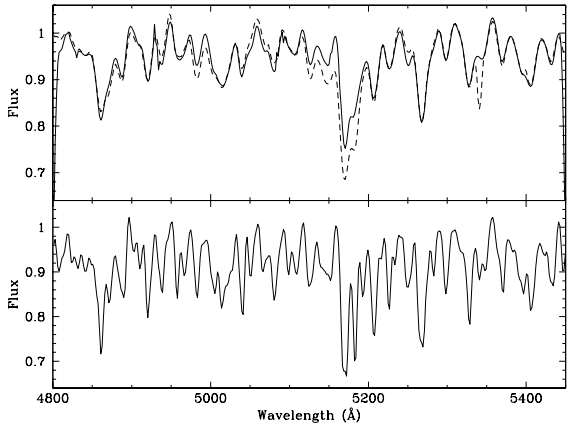}
\caption{Spectrum of the combined, weighted template stars (lower panel), data from the central bin along the minor axis (dashed line, upper panel), and the template spectrum convolved with the best-fitted LOSVD (solid line, upper panel).  The region from 5163 \AA\ to 5228 \AA\ is excluded from the fit.
\label{fig:spec}}
\end{figure}

For our kinematic analysis we use the spectral range 4800-5450\AA\
which matches the wavelength range of our template stars.  This region
includes the H$\beta$ and Mg$b$ lines, however we exclude the Mg$b$
region because it is enhanced \citep{pro05} and our template stars do
not provide a proper fit.  \citet{bar02} show that in pixel-space
fitting routines the Mg$b$ line is sensitive to template mismatch and
the details of the fitting procedure.  If the Mg$b$ line is included
in the fit, the measured dispersions are falsely high by as much as
20\% to account for the abundance discrepancy.  An example fit is
shown in Figure \ref{fig:spec}.

The uncertainty of each velocity bin is obtained from Monte Carlo
simulations.  We convolve the best-fitted LOSVD and weighted stellar
templates to obtain an initial galaxy spectrum.  We then generate 100
realizations of the galaxy spectrum by adding Gaussian noise using an
estimate of the initial rms.  The LOSVD is determined for each
realization as described above.  The distribution of values in each
velocity bin of the LOSVD provides an estimate of the 68\% confidence
bands.  The median value of the dispersion from the 100 realizations
compared to the initial dispersion reveals any possible bias in the
dispersion measurement.

\begin{figure}
\epsscale{0.8}
\plotone{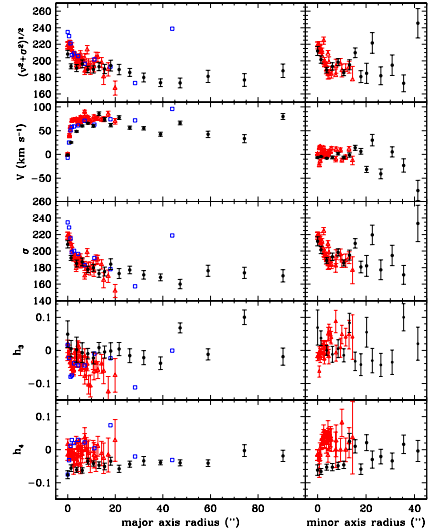}
\caption{Gauss-Hermite moments of the LOSVDs and the rms line-of-sight velocity: mean velocity $V$, velocity dispersion $\sigma$, asymmetric deviations from Gaussian (skewness) $h_3$, and symmetric deviations from Gaussian (kurtosis) $h_4$ along the major axis (left panel) and minor axis (right panel) for our data (black filled circles), SAURON \citep{ems04} (blue open triangles), and \citet{pin03} (red open squares).
\label{fig:pallmcall}}
\end{figure}

Although we use the full nonparametric velocity profile in the dynamic
modeling it is useful to compare moments of the distribution. In
Figure \ref{fig:pallmcall} we plot, from top to bottom, the second
moment as measured by $\sqrt(V^2+\sigma^2)$, the first four Gauss-Hermite
moments (mean velocity $V$, velocity dispersion $\sigma$, asymmetric
deviations from Gaussian (similar to skewness) $h_3$, and symmetric
deviations from Gaussian (similar to kurtosis) $h_4$).  The kinematic
data are given in Table \ref{tab:kinmj} and Table \ref{tab:kinmn}.  For
comparison, in Figure \ref{fig:pallmcall} we also plot data from
\citet{pin03} and \citet{ems04} extracted in a 1 arcsec slit along the
major and minor axes. The second moment of the line-of-sight velocity,
$(V^2+\sigma^2)^{1/2}$, is slightly smaller than the other samples
throughout the overlapping region.  This may be caused by a slit
misalignment (since $V$ will be higher on the major axis) or template
fitting difference. Since, however, our models are dominated by SAURON
data in the center, this difference is not a major issue. We further
run the dynamical models with using SAURON data alone and HET data
alone, and find un-biased results from when using the combined
dataset. 

\begin{deluxetable*}{ccccccccc}
\tablewidth{0pt}
\tablecaption{Major Axis Kinematics\label{tab:kinmj}}
\tablehead{
\colhead{$r$} & \colhead{$v$} & \colhead{$\epsilon_v$} & \colhead{$\sigma$} & \colhead{$\epsilon_{\sigma}$} & \colhead{$h_3$} & \colhead{$\epsilon_{h_3}$} & \colhead{$h_4$} & \colhead{$\epsilon_{h_4}$} \\ 
 $arcsec$ & $km~s^{-1}$ & & $km~s^{-1}$ & & & & &}
  \startdata  
  0.00 & -1.05 & 1.65 & 208.24 & 4.43 &  0.049 & 0.040 & -0.076 & 0.012 \\
  1.41 & 25.35 & 1.40 & 191.71 & 2.96 &  0.005 & 0.025 & -0.056 & 0.010 \\
  3.76 & 48.53 & 1.55 & 185.06 & 4.16 &  0.002 & 0.038 & -0.061 & 0.008 \\
  6.11 & 60.13 & 1.89 & 186.95 & 4.73 & -0.007 & 0.030 & -0.063 & 0.008 \\
  8.46 & 65.68 & 1.72 & 177.81 & 3.93 & -0.010 & 0.016 & -0.035 & 0.009 \\
 10.81 & 60.16 & 2.79 & 179.86 & 4.20 & -0.023 & 0.018 & -0.042 & 0.008 \\
 13.16 & 85.90 & 3.80 & 172.81 & 4.40 &  0.008 & 0.016 & -0.047 & 0.010 \\
 15.51 & 73.79 & 3.76 & 175.29 & 5.35 & -0.005 & 0.018 & -0.051 & 0.010 \\
 18.09 & 61.67 & 3.37 & 184.25 & 6.40 & -0.002 & 0.025 & -0.035 & 0.012 \\
 21.62 & 77.70 & 3.91 & 172.87 & 5.93 &  0.004 & 0.020 & -0.058 & 0.011 \\
 26.08 & 56.48 & 3.27 & 177.18 & 4.56 & -0.016 & 0.021 & -0.044 & 0.009 \\
 31.96 & 55.20 & 3.46 & 171.32 & 5.12 & -0.022 & 0.021 & -0.034 & 0.013 \\
 39.01 & 42.58 & 4.29 & 168.27 & 4.21 & -0.039 & 0.019 & -0.039 & 0.007 \\
 47.24 & 66.29 & 3.89 & 160.18 & 5.64 &  0.068 & 0.015 & -0.040 & 0.009 \\
 58.99 & 42.41 & 6.68 & 176.27 & 6.69 & -0.012 & 0.017 & -0.041 & 0.010 \\
 74.26 & 33.40 & 8.50 & 173.74 & 6.79 &  0.100 & 0.022 & -0.003 & 0.018 \\
 90.47 & 79.97 & 6.16 & 170.17 & 7.48 & -0.019 & 0.026 & -0.019 & 0.017 \\
 \enddata
 \end{deluxetable*}

\begin{deluxetable*}{ccccccccc}
\tablewidth{0pt}
\tablecaption{Minor Axis Kinematics\label{tab:kinmn}}
\tablehead{
\colhead{$r$} & \colhead{$v$} & \colhead{$\epsilon_v$} & \colhead{$\sigma$} & \colhead{$\epsilon_{\sigma}$} & \colhead{$h_3$} & \colhead{$\epsilon_{h_3}$} & \colhead{$h_4$} & \colhead{$\epsilon_{h_4}$} \\
 $arcsec$ & $km~s^{-1}$ & & $km~s^{-1}$ & & & & &}
  \startdata
  0.00 &  -6.51 &  1.29 & 211.91 &  5.50 &  0.069 & 0.053 & -0.062 & 0.016 \\
  1.41 &  -4.68 &  0.86 & 201.49 &  4.47 &  0.037 & 0.022 & -0.066 & 0.009 \\
  3.76 &  -7.94 &  1.04 & 188.69 &  3.30 &  0.000 & 0.015 & -0.053 & 0.009 \\
  6.11 &  -6.27 &  2.88 & 193.16 &  4.54 & -0.012 & 0.024 & -0.053 & 0.011 \\
  8.46 &   1.79 &  2.63 & 198.46 &  4.27 &  0.005 & 0.027 & -0.049 & 0.011 \\
 10.81 &  -6.24 &  2.72 & 185.65 &  3.66 & -0.015 & 0.020 & -0.046 & 0.011 \\
 13.16 &  -1.07 &  4.47 & 195.26 &  6.04 &  0.083 & 0.029 &  0.025 & 0.024 \\
 15.51 &  14.06 &  5.22 & 209.36 &  5.76 & -0.033 & 0.028 &  0.010 & 0.020 \\
 17.86 &   6.49 &  6.04 & 180.82 &  7.07 & -0.043 & 0.033 & -0.059 & 0.016 \\
 20.21 & -31.59 &  6.41 & 182.22 & 12.62 &  0.055 & 0.045 &  0.021 & 0.024 \\
 22.56 &  30.06 & 12.08 & 219.63 & 11.81 & -0.031 & 0.037 & -0.030 & 0.026 \\
 26.08 & -40.85 & 10.30 & 177.36 & 11.96 & -0.044 & 0.035 & -0.021 & 0.022 \\
 30.78 &   5.47 & 10.77 & 194.77 & 12.27 & -0.036 & 0.037 & -0.034 & 0.024 \\
 35.49 & -23.16 & 14.05 & 171.31 & 11.39 &  0.100 & 0.041 &  0.016 & 0.028 \\
 41.36 & -76.04 & 21.58 & 233.45 & 21.95 &  0.020 & 0.050 & -0.004 & 0.032 \\
 \enddata
 \end{deluxetable*}

Our $h_4$ values are more negative than the data from the literature.
This discrepancy could be attributed to template mismatch (either in
the published analysis or in ours), however we use a wide range of
template stars and do not get a different result when more template
stars are made available for the fit.  It could also be that relying
on Gauss-Hermite parameterization causes some differences since there
are known correlations, especially with higher order moments
\citep[see][]{mag06, hou06}. Since we fit the LOSVD directly in the
dynamical models, a better comparison would be with those profiles, as
opposed to their moments. The dark halo mass, however, is determined
mainly by the radial profile of the second moment, and $h_4$
determines mainly the anisotropies. There is certainly some degeneracy
between the two parameters, but we find no reason to believe that our
$h_4$ values are incorrect. There are also kinematic points that have
differences which are inconsistent with their reported uncertainties
(for example, some of the minor axis points), and the uncertainties
may be underestimated for those points. We run halo models without the
most discrepant points and still find the same halo results as when
they are included.

\citet{pro09} measure the kinematics of NGC~821 using the background galaxy light from Keck DEIMOS multi-object spectroscopy of globular clusters.  Their results show good agreement with our $V$ and $\sigma$ profiles, as seen in their Figure~17.  Previously, \citet{pro05} determined the kinematics of NGC~821 using the Gemini Multi-Object Spectrograph (GMOS).  As shown in Figure~2 of that paper, our $\sigma$ is in agreement with theirs, except that they find a lower dispersion at about $30\arcsec$.  However their data are of substantially lower signal to noise and they are also unable to determine higher order moments.

\citet{wei09} provide a new analysis of the SAURON data and
also include additional data at large radii. Their furthest radial
point is at 110\arcsec\ (which they refer to as 4 Re), whereas our
last point is at 90\arcsec\ (which we refer to as 2 Re).  The
comparison between the two kinematic sets is shown in Figure~6 from
\citet{wei09}. There is generally excellent agreement between the two
sets of kinematics. Furthermore, their re-analysis of the SAURON
central pointing shows $h_4$ values now more consistent with our numbers.
The higher-order moments of the LOSVD are difficult to measure, and it
is important to consider systematic difference in the analysis. The
spectra from \citet{wei09} have lower signal-to-noise than our
spectra, which could add to systematic difference. For this reason,
our dynamical modeling does not include their kinematics, although we
suspect there will be little difference in the overall results. Thus,
we can compare the constraints on the dark matter parameters, which
would include systematic differences in the kinematic samples used.

Our results are consistent with \citet{wei09}, and not consistent
with either \citet{rom03} or \citet{coc09}. The Weijmans et
al. data are consistent with that of Coccato et al. The reason is
simply that the Weijmans et al. dispersions have large uncertainties,
and that the Weijmans et al. dispersions are between our values and
that of Coccato et al. The difference is not due to comparing
dispersions measured from major axis radii compared to circular radii
(as in Coccato et al.), since our dispersion profile is nearly
flat. We compare with the Romanowsky et al. values and not to the
values reported in Coccato. The differences between Coccato et al. and
Romanowsky et al. are small enough to not impact our analysis or
conclusions. The Coccato et al. data have slightly smaller
uncertainties than Romanowsky et al., so the statistical differences
between our dispersions and theirs is slightly larger.

\begin{figure}
\includegraphics[angle=270,scale=.3]{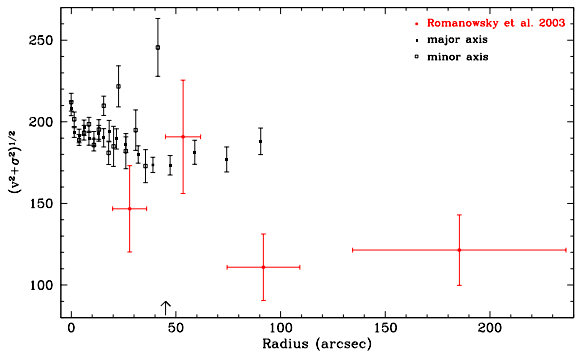}
\caption{The rms line-of-sight velocity $(v^2+\sigma^2)^{1/2}$ in km s$^{-1}$ as a function of radius from our major axis data (filled squares) and minor axis data (open squares), compared to the data from planetary nebulae measurements \citep{rom03}.  We use $v$ and $\sigma$ as measure from a Gauss-Hermite fit; since we do not correct for the higher order moments, these values approximate the actual second moment.  The arrow indicates the adopted $R_e$ of the galaxy.  
\label{fig:pallmcpn}}
\end{figure}

Figure \ref{fig:pallmcpn} shows the rms line-of-sight velocity
$(V^2+\sigma^2)^{1/2}$ compared to the planetary nebula results of
\citet{rom03}.  Our largest radii data show higher rms line-of-sight
velocities than the planetary nebulae, at about $3\sigma$ for their
two largest radii points. Thus, there appears to be a significant
difference in the kinematics between the two samples.

\section{Dynamical Models}\label{mod}

We use axisymmetric orbit superposition models based on the method of
\citet{sch79}. The surface brightness profile is converted to a
luminosity density profile using an assumed inclination.  We assume an
edge-on inclination for this analysis, which is reasonable given
NGC~821's large ellipticity of $0.40$ \citep{cap07}.  This luminosity
density is converted to a mass density using a mass-to-light ratio
($M/L_V$) that is constant over the galaxy.  A spherically symmetric
dark halo density profile is added to the stellar density and this
total mass density gives the galaxy's gravitational potential.  Next
individual stellar orbits are sampled in energy ($E$), angular
momentum ($L_z$) and the third integral ($I_3$) and these orbits are
integrated in the specified potentials.  The galaxy is divided
spatially into cells both in real space and in projection, and the
amount of time that an orbit spends in a cell represents the mass
contributed by that orbit.  The orbits are combined with nonnegative
weights to find the best-fitted superposition to match the data LOSVDs
from both HET and SAURON and the light profile.  The model incorporates seeing by convolving the light 
distribution for every orbit with the appropriate PSF before comparing with the data. This process is
repeated for different dark halo density profiles and $M/L_V$ values
to find the halo potential that best fits the data, as determined by
$\chi^2$ (described in \S \ref{res}).

The SAURON data are described in \citet{ems04}.  We reconstruct a full LOSVD from their reported moments and rebin them to match our model bins.  The \citet{pin03} data shown in Figure \ref{fig:pallmcall} is not used in the models.

To reduce computational time, an orbit library is calculated for a
given input dark halo plus stars with a mass-to-light ratio of one.
The velocities are then scaled accordingly given the mass-to-light
ratio before matching the data. The numbers reported are the actual
density parameters, including this $M/L$ factor, which gives the
somewhat irregular parameter space grids (as seen in Figure
\ref{fig:2dchinfw}).

We use the orbital weight fitting of \citet{geb00,geb03} with the
orbit library sampling of \citet{tho04,tho05}.  Our models differ from
others \citep[e.g.][]{cre99} in that we use maximum entropy
\citep{ric88} and we utilize the full LOSVD, rather than its moments.
\citet{tho04,tho05} show the ability of our orbit libraries to
recovery dark halo profile from mock elliptical galaxy data.
Therefore these models should accurately measure the properties of
NGC~821 given the caveats that we assume an axisymmetric galaxy and
spherical halo.

The orbits are computed in 4 angular bins and 15 radial bins from
0.3$\arcsec$ to 300$\arcsec$.  These bins are similar in size to the HET
data extraction bins in the radial direction, though in the angular direction they span about 20 degrees. We specify the galaxy potential and the forces on a grid 
that is four times finer.  Our libraries have approximately 10000 total
orbits.

To calculate our galaxy potential we use a composite surface
brightness profile.  Within $0.3\arcsec$ we use the profile from
\citet{lau05} as compiled in \citet{pin03} based on HST WFPC2 images
in F555W.  Outside of $0.3\arcsec$ we use a composite profile from HST
PC F555W and the McDonald Observatory 0.8-m telescope in $V$
(D. Fisher, private communication).  The surface brightness
deprojection is based on a nonparametric estimate of the density using
smoothing splines \citep[see][]{geb96}.  The luminosity density is given in Table \ref{tab:lumdens}.

\begin{deluxetable}{cc}
\tablewidth{0pt}
\tablecaption{V-band Luminosity Density\label{tab:lumdens}}
\tablehead{
\colhead{radius (\arcsec)} & \colhead{$L_{\odot}/pc^3$} } 
 \startdata
 2.300E-02 & 8.275E+03 \\
 2.533E-02 & 7.205E+03 \\ 
 2.790E-02 & 6.283E+03  \\
 3.073E-02 & 5.485E+03  \\
 3.385E-02 & 4.793E+03  \\
 3.728E-02 & 4.192E+03  \\
 4.107E-02 & 3.668E+03  \\
 4.523E-02 & 3.212E+03  \\
 4.982E-02 & 2.813E+03  \\
 5.487E-02 & 2.464E+03 \\
 \nodata & \nodata \\
 \enddata
 \end{deluxetable}

\subsection{NFW Halo}\label{nfwhalo1}

We use the NFW \citep{nfw96} dark halo density profile, given as
\begin{equation}
\rho(r)=\frac{\rho_{crit}\ \delta_c}{(r/r_s)(1+r/r_s)^2}
\end{equation}
where $r_s$ is the scale radius of the halo, $\rho_{crit}=3H^2/8\pi
G$ is the critical density, and $\delta_c$ is the characteristic overdensity.  We use $H=70$ km s$^{-1}$ Mpc$^{-1}$.
Throughout this paper we refer to $\rho_{crit}\delta_c$ as the scale
density.  We use two independent parameters to define the NFW halo in our moedls: the scale density and the scale radius.  The scale density can also be written in terms of a concentration parameter $c$ by
\begin{equation}
\delta_c=\frac{\Delta_{vir}}{3}\frac{c^{3}}{\ln{(1+c)}-c/(1+c)}.
\end{equation}
The virial overdensity $\Delta_{vir}$ varies with redshift and
cosmological model and we use a value of $\Delta_{vir}=101$.  Although we vary
both the concentration (density) and scale radius, there is a known
correlation between them \citep{nfw96}.  This relation as given in
\citet{bul01} is
\begin{equation}
c\simeq9\left(\frac{M_{vir}}{1.5\times10^{13}h^{-1}M_{\sun}}\right)^{-0.13}
\label{eqn:nfwMvir}
\end{equation}
and can be written in the form
\begin{equation}
r_{s}^{3}=\left(\frac{c}{9}\right)^{-1/0.13}\left(\Delta_{vir}\frac{4\pi}{3}\rho_{crit}c^3\right)^{-1}(1.5\times10^{13}h^{-1}M_{\odot}).
\label{eqn:nfwcr}
\end{equation}

\subsection{Power-Law Halo}\label{powhalo1}

The best-fitted NFW halo profiles have a break radius beyond the
extent of our modeling and therefore look like a power-law over the
extent of our models (see \S \ref{nfwhalo} below).  We therefore tried
a simple power-law profile as well.  We used power-law density
profiles of the form
\begin{equation}
\rho(r)=\rho_o(\frac{r}{r_o})^{-n}
\end{equation}
where $n$ is the power-law slope, $\rho_o$ is the characteristic
density, and $r_o$ is the characteristic radius such that
$\rho(r=r_o)=\rho_o$.  We use $r_o=0.3\arcsec=34\ pc$ because it is
the inner-most radial point calculated in the models.

\section{Results}\label{res}

\begin{figure}
\includegraphics[angle=270,scale=.3]{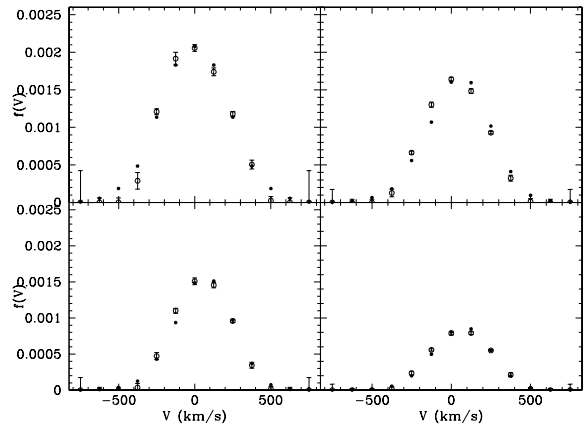}
\caption{Match of data and no dark halo model LOSVDs for the central four radial bins along the major axis (r = 0.00, 1.41, 3.76, 6.11 arcsec).  The open circles are the data values with error bars and the closed circles are the model values.  The area is normalized to the total light in that bin.
\label{fig:losvdm}}
\end{figure}

The best-fitted model is determined by comparing the $\chi^2$ between
the model and data LOSVDs, with the uncertainty of the data determined
from the 68\% confidence band.  Example LOSVDs are shown in Figure
\ref{fig:losvdm} for several radial bins. The measure of the reduced
$\chi^2$ is not straight-forward since because determining the number
of degrees of freedom is uncertain. The number of independent
observables is roughly the number of radial data bins times the number
of LOSVD bins at each radius ($69\times13=897$ in this case), however
the LOSVD bins are correlated and thus the effective number of data
points is less than this value. The best-fitted model has a $\chi^2$
of around 2200, and with 897 data points, this provides a large
reduced $\chi^2$. Typical values of the reduced $\chi^2$ for the
orbit-based models are around 0.5 (see Gebhardt et al. 2003), so the
value reported here is not typical. The main driver for the large
$\chi^2$ is the minor axis data---removing this data gives a reduced
$\chi^2$ below one. Furthermore, the results on the parameters do not
change significantly. Regardless, the change in $\chi^2$ between
different models remains a valid statistic to determine confidence
levels of the fits.  For example, a change in $\chi^2$ of 2.3
corresponds to the 68.3\% confidence level because we marginalize over
$M/L$ and thus have two parameters describing the halo.

Because of computational limits we first calculate models using a
coarse grid of mass-to-light ratio.  The $\chi^2$ values are then fit
with the IDL quadratic interpolation routine, and those models with
the lowest minimum $\chi^2$ are modeled with a finer mass-to-light
ratio interval. 

\subsection{NFW Halo}\label{nfwhalo}

\begin{figure}
\includegraphics[angle=270,scale=0.4]{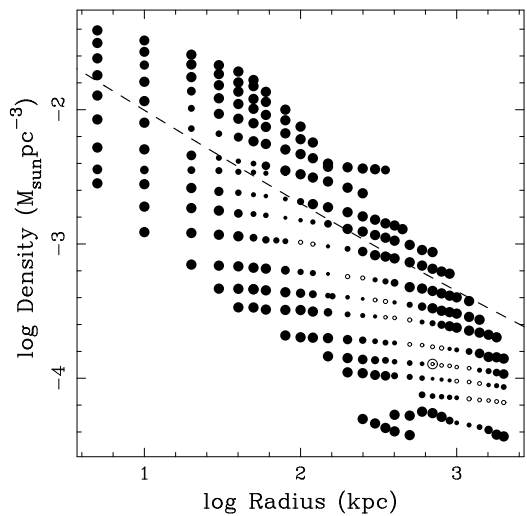}
\caption{Scale radius ($r_s$) and scale density ($\rho_{crit}\delta_c$) $\chi^{2}$ grid for NFW halo density profiles.  Each point represents a model, and the size of the point reflects the value of $\Delta\chi^2$ for the best-fitted $M/L_V$ value.  Models with $\Delta \chi^2$ less than $6 \sigma$ from the minimum value are plotted with open circles.  The ringed point indicates the model with the lowest value of $\chi^2$.  The dashed line shows the expected NFW parameter relation (see \S\ref{nfwhalo1}).
\label{fig:2dchinfw}}
\end{figure}

\begin{figure}
\plotone{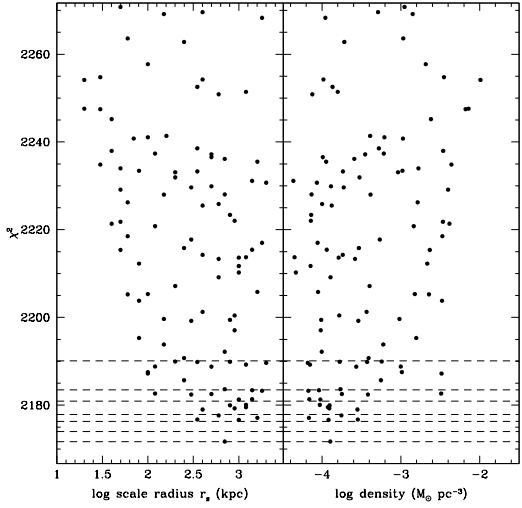}
\caption{$\chi^2$ as a function of scale radius and scale density ($\rho_{crit}\delta_c$) for NFW halo density profiles with best-fitted $M/L_V$.  The points represent actual modeled values.  The dashed lines refer to $\Delta\chi^2=2.3, 4.61, 6.17, 9.21, 11.8, 18.4$, corresponding to 2 degree of freedom confidence levels of 63.8\%, 90\%, 95.4\%, 99\%, 99.73\%, and 99.99\%.
\label{fig:2dchia4nfw}}
\end{figure}

We use models with scale radius from 1 to 2000 kpc and scale density
from $0.05$ to $3.0\times10^{-5} M_\sun pc^{-3}$, corresponding to a
range in c of approximately 0.75 to 23, and $M/L_V$ from 1.0 to 9.0.
Figures \ref{fig:2dchinfw} and \ref{fig:2dchia4nfw} show the resulting
$\chi^2$ as a function of halo scale radius and scale density.  The
points represent actual modeled values, and the $M/L_V$ that gives the
lowest $\chi^2$ is used at each point.  The dashed line in Figure
\ref{fig:2dchinfw} indicates the expected correlation of concentration
and scale radius as described in \S \ref{nfwhalo1}.  This relation has
a scatter of $\Delta \log r_s = 0.36$ \citep{bul01}.  Our data show a
degeneracy between scale radius and scale density that is similar to,
though slightly tilted from, the correlation.

$\chi^2$ is a function of 3 variables: stellar M/L, dark halo scale
radius and dark halo normalization. Due to computer resources, we do
not provide a uniformally-sampled grid of the 3 variables for the
$\chi^2$. Because of this, it is difficult to produce reliable
contours for any 2 of the parameters. Figure \ref{fig:2dchinfw} thus
shows only the location of the points (with size related to
$\chi^2$). We do not estimate uncertainties from the contours
directly, but instead rely on plotting $\chi^2$ versus each of the
parameters, including all values for the other two parameters. Figure
\ref{fig:2dchia4nfw} shows $\chi^2$ versus scale radius and
density. Uncertainties come from the envelope of these one-dimensional
plots. Since we have explored neither a regular grid nor a full set of
variables (e.g., black hole mass, inclination, change in the stellar
M/L with radius), the uncertainties should only be used in a
comparative sense with the models that we have tried.  A full
exploration of the uncertainties will come as computer resources
improve.

\begin{figure}
\plotone{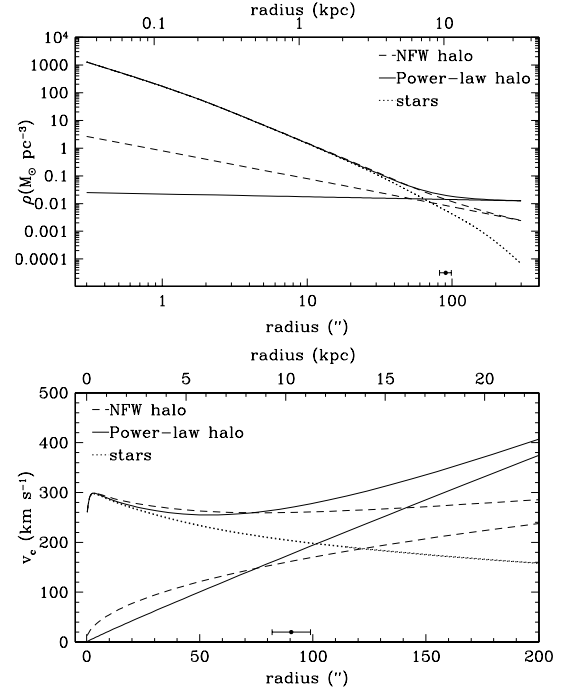}
\caption{Density (top) and circular velocity (bottom) as a function of radius for the best-fitted NFW (dashed lines) and power-law (solid lines) dark halos.  In each case the bottom line is the dark halo alone and the top line is the total mass (halo plus stars)  The data point shows the radius of the most extended bin of our kinematic data.
\label{fig:densvrpownfw}}
\end{figure}

We find that the best-fitted NFW dark halo density profile has scale
radius $700^{+500}_{-300}$ kpc and scale density
$1.28^{+0.8}_{-0.5}\times10^{-4} M_{\odot}pc^{-3}$, corresponding to a
c of 2.45.  The no halo model is ruled out with a change in $\chi^2$
of 356 (greater than 99\% confidence level) from the best-fitted NFW
halo.  Table \ref{tab:modres} shows the $\chi^2$ values and halo
parameters of the best-fitted halo model and model with no dark halo.
We do not attach significance to this density, radius, and
concentration.  They are clearly outside the expected range for a
galaxy and merely indicate that the NFW profile is not reasonable.
The halo needs more mass at large radii to fit the data.  The
best-fitted NFW halo density profile is shown in Figure
\ref{fig:densvrpownfw}.  The scale radius is well beyond the radial
extent of our modeling, and is indicative of the need for a near
power-law profile over the extent of our models.

\begin{figure}
\includegraphics[angle=270,scale=0.3]{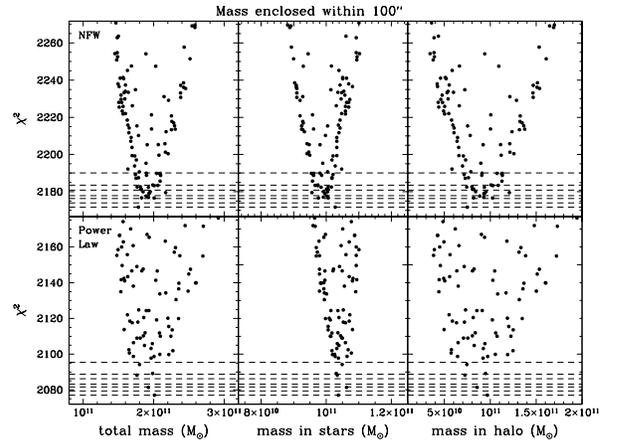}
\caption{Enclosed mass within the radial extent of our kinematic data, $100\arcsec$, as a function of $\chi^2$ for both the NFW halo density profiles (top) and power-law halo density profiles (bottom) with best-fitted $M/L_V$.  The dashed lines refer to $\Delta\chi^2=2.3, 4.61, 6.17, 9.21, 11.8, 18.4$, corresponding to 2 degree of freedom confidence levels of 63.8\%, 90\%, 95.4\%, 99\%, 99.73\%, and 99.99\%.  
\label{fig:totmass}}
\end{figure}

Since dynamical modeling directly measures mass (as opposed to dark
halo parameters), the enclosed mass provides a more robust estimate
and is likely not subject to the specific parameterization of the dark
halo. Figure \ref{fig:totmass} shows the mass enclosed within the
extent of our kinematic data as a function of $\chi^2$.  The
best-fitted total enclosed mass is $1.78\pm0.15\times10^{11}M_{\sun}$,
divided into $1.03\pm0.03\times10^{11}M_{\sun}$ in stars and
$0.75\pm0.15\times10^{11}M_{\sun}$ in dark matter.  At $1R_e$ the
ratio of dark matter to total matter is 0.19.  The best-fitted NFW
halo circular velocity profile is shown in Figure
\ref{fig:densvrpownfw}.
 
\begin{figure}
\includegraphics[angle=270,scale=0.3]{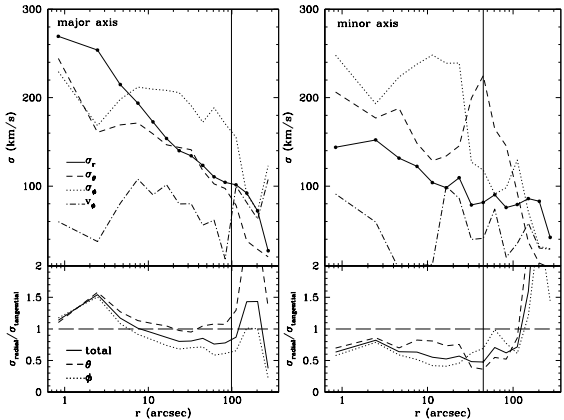}
\caption{Internal moments $\sigma_r$, $\sigma_{\theta}$, and $\sigma_{\phi}$ (top) and the ratio of radial to tangential dispersion (bottom) along the major axis (left) and minor axis (right) for the model with no dark halo.  Note that $\sigma_{\phi}$ includes both random and ordered motions, which are shown (dot-dashed line) and are small.  The vertical line shows the limit of our kinematic data; results beyond this radius are not reliable.
\label{fig:intmom000}}
\end{figure}

\begin{figure}
\includegraphics[angle=270,scale=0.3]{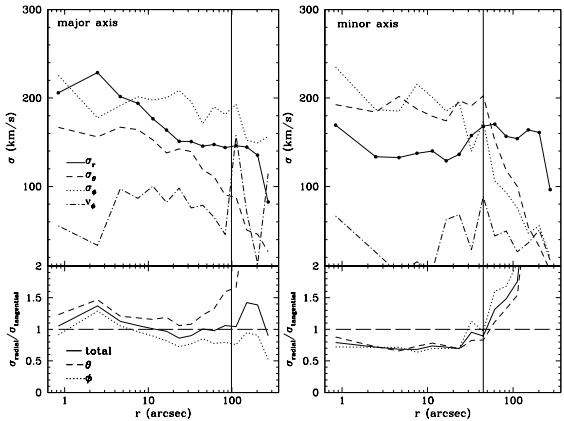}
\caption{Internal moments $\sigma_r$, $\sigma_{\theta}$, and $\sigma_{\phi}$ (top) and the ratio of radial to tangential dispersion (bottom) along the major axis (left) and minor axis (right) for the model with the best-fitted NFW halo.  Note that $\sigma_{\phi}$ includes both random and ordered motions, which are shown (dot-dashed line) and are small.  The vertical line shows the limit of our kinematic data; results beyond this radius are not reliable.
\label{fig:intmom256}}
\end{figure}

Figures \ref{fig:intmom000} and \ref{fig:intmom256} show the internal
moments $\sigma_r$, $\sigma_{\theta}$, and $\sigma_{\phi}$ and ratio
of radial to tangential dispersion along the major and minor axes for
the model with no dark halo and the best-fitted NFW halo model. The
model without a dark halo shows radial anisotropy at small radii and
tangential anisotropy at large radii along the major axis.  Tangential
anisotropy at large radii in a model with no dark halo could be an
indication of the need for a dark halo because the observations
largely constrain only $\sigma_{\phi}$ (for an edge-on configuration),
so both $\sigma_r$ and $\sigma_{\theta}$ may be artifically decreased
to create a smaller total $\sigma$ that can be fit without a dark
halo.  Along the minor axis, the contribution in the $\theta$ and
$\phi$ directions are roughly equal, as is expected for an
axisymmetric model.  Overall the minor axis shows tangential
anisotropy over the entire range of our data.  The NFW model is more isotropic than the model with no halo.

\begin{figure}
\plotone{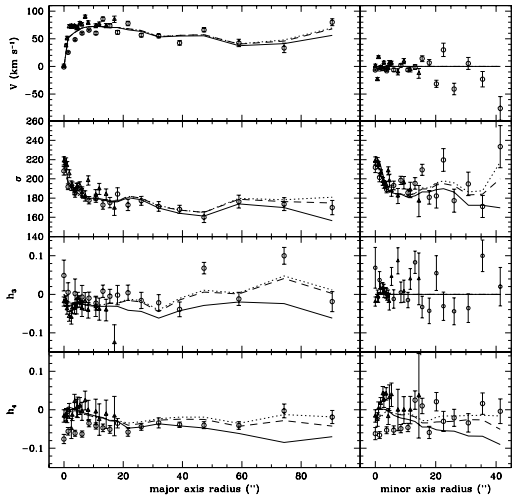}
\caption{Gauss-Hermite moments (mean velocity $V$, velocity dispersion $\sigma$, asymmetric deviations from Gaussian (skewness) $h_3$, and symmetric deviations from Gaussian (kurtosis) $h_4$) of the LOSVDs for our HET data and the best-fitted halo models along the major axis (left panel) and minor axis (right panel).  The HET  data are shown with open circles, SAURON data along the axes with open triangles, no dark halo model with solid lines, best-fitted NFW halo with dashed lines, and best-fitted power-law halo with dotted lines.  The model fits the full LOSVD of the HET data and all of the SAURON data.  
\label{fig:gherm}}
\end{figure}

Although the models fit the full nonparametric velocity profile of
both the HET data and SAURON data, in Figure \ref{fig:gherm} we plot
the first four Gauss-Hermite moments for our HET data and the
best-fitted halo models.  The models differ most at intermediate to
large radii, and do not appear to be driven by any one single parameter or
radius in particular.

\subsection{Power-Law Halo}\label{powhalo}

\begin{figure}
\includegraphics[angle=270,scale=0.4]{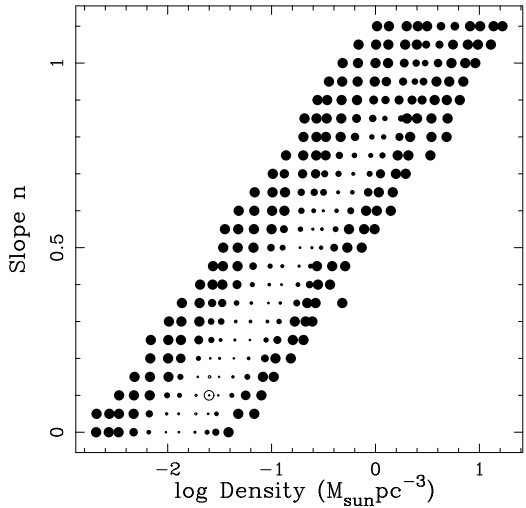}
\caption{Power-law slope n and density $\rho_o$ $\chi^{2}$ grid for power-law halo density profiles.  Each point represents a model, and the size of the point reflects the value of $\Delta\chi^2$ for the best-fitted $M/L_V$ value.  Models with $\Delta \chi^2$ less than $6 \sigma$ from the minimum value are plotted with open circles.  The ringed point indicates the model with the lowest value of $\chi^2$.  \label{fig:densrchipow}}
\end{figure}

\begin{figure}
\plotone{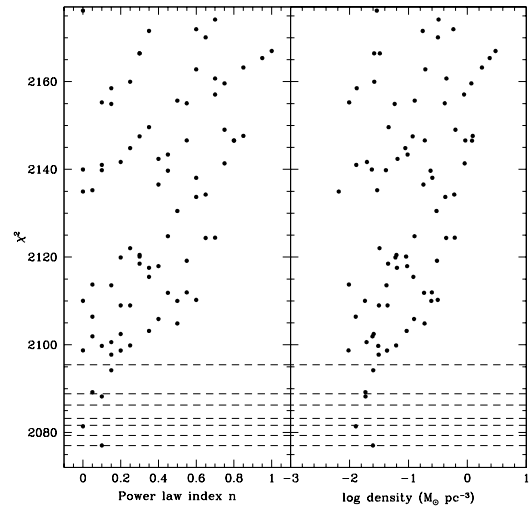}
\caption{$\chi^2$ as a function of power-law index $n$ and scale density $\rho_o$ for power-law halo density profiles with best-fitted $M/L_V$.  The points represent actual modeled values.  The dashed lines refer to $\Delta\chi^2=2.3, 4.61, 6.17, 9.21, 11.8, 18.4$, corresponding to 2 degree of freedom confidence levels of 63.8\%, 90\%, 95.4\%, 99\%, 99.73\%, and 99.99\%.
\label{fig:chiparampow}}
\end{figure}

We run models with a range of slope $n$ from 0.0 to 1.1, density
$\rho_o$ from 0.0015 to 26 $M_{\odot}/pc^3$, and $M/L_V$ from 3.5 to
8.0.  The resulting $\chi^2$ grid is shown in Figure
\ref{fig:densrchipow} and as a function of $n$ and $\rho_o$ in Figure
\ref{fig:chiparampow}.  The best-fitted halo model has a slope
$0.1^{+0.1}_{-0.08}$ and a characteristic density
$\rho_o=0.025^{+0.025}_{-0.009} M_{\odot}pc^{-3}$.  This halo is a
better fit to the data than the best NFW halo, with a
$\Delta\chi^2=95$ (see Table \ref{tab:modres}).  This power-law slope
is significantly more shallow than the 1.0 slope of an NFW profile.  A
comparison of the best-fitted halo density and circular velocity
profiles is shown in Figure \ref{fig:densvrpownfw}.

\begin{deluxetable*}{ccccccccccc}
\tabletypesize{\footnotesize}
\tablewidth{0pt}
\tablecaption{Best-Fitted Halo Model Results\label{tab:modres}}
\tablehead{
\colhead{halo} & \colhead{$\chi^2$} & \colhead{$M/L_V$} & \colhead{$r_s$} & \colhead{$\rho$} & \colhead{$c$} & \colhead{$M_{vir}$} & \colhead{$n$}  & \colhead{$M_{tot}$} & \colhead{$M_{stars}$} & \colhead{$M_{halo}$}\\
 & & ($M/L)_{\sun}$ & $kpc$ & $M_{\odot}/pc^3$ & & $10^{17}M_{\odot}$ & & $10^{11}M_{\odot}$ & $10^{11}M_{\odot}$ & $10^{11}M_{\odot}$\\
(1) & (2) & (3) & (4) & (5) & (6) & (7) & (8) & (9) & (10)}
 \startdata
none & 2527.30 & $7.25\pm0.05$ & \nodata  & \nodata  & \nodata & \nodata  & \nodata & $1.20\pm0.01$ & $1.20\pm0.01$ & 0.00  \\
NFW & 2171.70 & $6.19\pm0.09$ & $700^{+500}_{-300}$ & $1.28^{+0.8}_{-0.5}\times10^{-4}$ & $2.45^{+0.65}_{-0.53}$ & $4.76^{+26}_{-4.0}$ & \nodata & $1.78\pm0.15$ & $1.03\pm0.03$ & $0.75\pm0.15$\\
power-law & 2077.05 & $6.25\pm0.07$ & \nodata  & $0.025^{+0.025}_{-0.009}$ & \nodata & \nodata & $0.1^{+0.1}_{-0.08}$ & $2.01\pm0.15$ & $1.04\pm0.02$ & $0.97\pm0.15$ \\
 \enddata
\tablecomments{(1) Dark halo density profile. (2) $\chi^2$ of best-fitted model. (3) Stellar $M/L_V$ of best-fitted model. (4) Scale radius of best-fitted model. (5) Scale density $\rho_{crit}\delta_c$ for NFW, characteristic density $\rho_o$ for power-law. (6) NFW concentration parameter determined from scale density. (7) Virial mass determined from NFW concentration parameter. (8) Power-law index. (9) Total mass within 100\arcsec. (10) Mass of stars within 100\arcsec. (11) Mass of dark halo within 100\arcsec.}
 \end{deluxetable*}

\begin{figure}
\includegraphics[angle=270,scale=0.3]{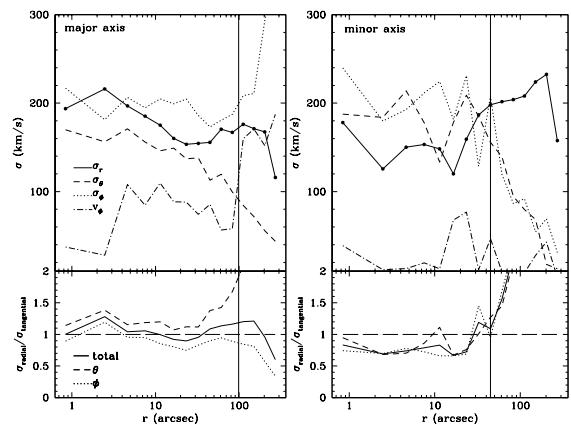}
\caption{As in Figure \ref{fig:intmom000} for the best-fitted power-law model.
\label{fig:intmom027}}
\end{figure}

The best-fitted total enclosed mass is
$2.01\pm0.15\times10^{11}M_{\sun}$, divided into
$1.04\pm0.02\times10^{11}M_{\sun}$ in stars and
$0.97\pm0.15\times10^{11}M_{\sun}$ in dark matter (see Figure
\ref{fig:totmass}).  At $1R_e$ the ratio of dark matter to total
matter is 0.13. The internal moments $\sigma_r$, $\sigma_{\theta}$,
and $\sigma_{\phi}$ and ratio of radial to tangential dispersion along
the major axis are shown in Figure \ref{fig:intmom027} and are roughly
consistent with those of the best-fitted NFW halo.

Figure \ref{fig:gherm} shows the first four Gauss-Hermite moments for
our HET data and the best-fitted halo models.  Note that the models
fit the full nonparametric velocity profile of both the HET data and
SAURON data.

\subsection{Model Tests}\label{tests}

\begin{figure}
\includegraphics[angle=270, scale=0.4]{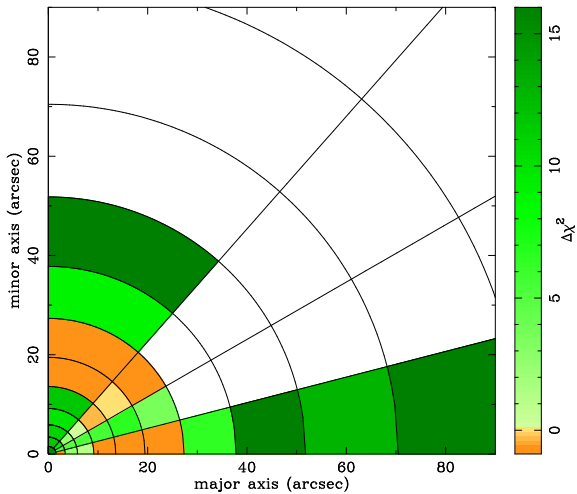}
\caption{Difference in $\chi^2$ between the LOSVDs of the model with no dark halo and the best-fitted power law halo model averaged in each spatial bin. Green indicates that the no-halo model has a larger $\chi^2$ than the power-law model and therefore the power-law is a better fit, while orange indicates that the power-law model has a larger $\chi^2$ than the no-halo model and therefore the no halo model is a better fit. 
\label{fig:bins}}
\end{figure}

In order to learn which aspect of the data is driving the results we
have performed several tests.  Using an abbreviated grid of about one
third of the halos in the full run and a coarse spacing in $M/L_V$ we
have re-run the models with various subsets of the data.  First, to
address any concerns over the scattered minor axis data we have run
the test models using only the HET major axis data and SAURON data.
The results are the same as the full data set; there is a clear need
for a dark halo, and the best-fitted NFW halo is not as good a fit as
the power-law halo.  Second, we removed the two points on the major
axis with extreme $h_3$ values (at about $47\arcsec$ and $74\arcsec$)
since Figure \ref{fig:gherm} may lead one to believe they are driving
the fits.  Again the results are the same as with the full data set.
And third, we do a test run using only data below $0.5 R_e$.  In this
case there is essentially no difference in $\chi^2$ between the three
best-fitted models (no dark halo, NFW halo, and power-law halo), and
the best-fitted halos are quite different than those from the full
data results.  These tests indicate that it is the large radii data as
a whole that is driving the model fits.  To further demonstrate this,
Figure \ref{fig:bins} shows the $\Delta\chi^2$ between the model with
no halo and the best-fitted halo in each bin.  The bins at large radii
show the greatest change in $\chi^2$, again indicating that the large
radii data are the major factor in the fits.

\subsection{Comparisons to Other Studies}\label{comparison}

\citet{geb03} model the central region of NGC~821 and find that it is
radially anisotropic within a few arcseconds and isotropic to slightly
tangentially anisotropic at larger radii.  Given the difference in
spatial resolution this roughly agrees with our result.  \citet{cap07}
find that within about $20 \arcsec$ N821 is radially anisotropic
overall, along the major and minor axes and between. They find that
the velocity ellipsoids are circular in the center and become more
radial with increasing radius, in conflict with our results, but they
do not include a dark halo which could change their results.
\citet{tho07}, using similar modeling as we use, find that early-type
galaxies in the Coma cluster are radial compared to the $\theta$
direction over all radii along the major axis, agreeing with our
result, and vary from galaxy to galaxy in the $\phi$ component.  The
merger simulations of \citet{dek05} also find a radial anisotropy.
Their spherically averaged $\beta$ of about $0.4$ corresponds to a
$\sigma_{radial}/\sigma_{tangential}$ of $1.3$, which is larger than
our results along the major and minor axes.  However their simulations
show declining projected dispersion profiles, which our data does not,
that could account for the difference.

By modeling only the central part of NGC~821 \citet{geb03} find
$M/L_V=7.6$ (without including foreground extinction), which is
consistent with our no-halo value over the whole galaxy of
$M/L_V=7.25$.  Correcting for NGC~821's large reddening of $A_V=0.364$
mag \citep[NED extragalactic database]{sch98} we find our best-fitted
$M/L_{V,no halo}=5.18$, $M/L_{V,nfw}=4.43$, and $M/L_{V,pow}=4.47$.
\citet{cap06} find $M/L_{jeans}=3.54$, $M/L_{schwarzschild}=3.08$, and
$M/L_{stellar pop}=2.60$ in the $I$ band.  Using $(V-I)=1.35$ mag
\citep{lau05}, dereddened to $(V-I)_0=1.20$ mag, and $(V-I)_{\sun}=0.682$ mag \citep{ram05} our $V$ band
mass-to-light ratios are converted to $M/L_{I,no halo}=3.21$,
$M/L_{I,nfw}=2.75$, and $M/L_{I,pow}=2.77$.  Our mass-to-light ratios
are slightly higher, though roughly consistent, with their
mass-to-light ratios found using Schwarzschild modeling and stellar
populations. 

We find that the enclosed mass of NGC~821 within $\sim 2 R_e$ is
roughly $2\times10^{11}M_{\sun}$, equally divided between stars and
dark matter.  At $1R_e$ the ratio of dark matter halo mass to total
mass is $0.19$ for the best-fitted NFW halo profile and $0.13$ for the
best-fitted power-law halo profile.  This matches other studies that
find that the dark matter is $10-40\%$ of the total matter at $1R_e$
and that dark matter begins to dominate at $2-4R_e$
\citep[e.g.][]{sag00,ger01,mam05}.  The simulations of \citet{dek05}
also show that dark matter and stellar matter are equal at $3 R_e$,
and at $1 R_e$ have a mass fraction of $40\%$ dark matter.
\citet{tho07} perform similar dynamical modeling on 17 galaxies in the
Coma cluster.  Using values taken by eye from their Figure 5 we find
that their average dark matter fraction at $1R_e$ is $0.19$, though
their galaxies show a wide range of dark matter fractions, from about
$0.1$ to $0.5$ at $1R_e$.  We therefore find that the dark matter
fraction at $1R_e$ is similar for N821, a field elliptical galaxy, and
a selection of Coma cluster early-type galaxies, perhaps contrary to
hypotheses that environment plays a role in the dark matter fraction.

\citet{wei09} provide a dynamical analysis using orbit-based
models and using data that extend to similar radii, though they do not attempt to characterize the shape of the halo. Thus, the
comparison of dark halo mass results is informative. We find very similar
numbers.  Inside of 39\arcsec\ (which they call Re), they find a dark
matter fraction of 18\%.  Inside of 45\arcsec\ (which we call Re), we
have a dark matter fraction of 13\%. There are differences in the
models as well. First, they use a ``maximum M/L'' model where they
force the M/L of the stars to have a maximum value. We find the best-fit stellar M/L amongst the range modeled. Second, the modelling codes are different, with
the main difference in that they use regularization (which trades the
best fitted values with smoothness) and we report results for the best
fit to the data. Third, they use SAURON data at large radii and we use
our HET data at large radii. Their data extend to slightly larger
radii (110\arcsec\ compared to our limit of 90\arcsec), and our data
is high signal-to-noise. Given all of the these differences, it is
impressive that we obtain similar results for the dark halo mass. This
implies that systematic differences are not significant for
determining the enclosed mass profile.

\subsection{Smoothing}\label{smooth}

It is useful to constrain the orbital weighting so that the resulting
DF is smooth, as a real galaxy's DF may be presumed to be.  Although
we do not usually report results when smoothing our models (we argue
that allowing the best fit to the data is the most robust way to
provide an un-biased result), other groups suggest that it is
important for their model. \citet{rix97} and subsequent studies
minimize the variation in the DF, a process they term regularization.
We employ maximum entropy to find the best combination of orbit
weights to match the data, as described in \citet{tho05}.  We define a
function $f\equiv\chi^2-\alpha S$ where $\chi^2$ is the sum of squared
residuals to the data, $S$ is the entropy, and $\alpha$ is a parameter
describing the relative weights of entropy and residuals in the fit.
In order to minimize $f$ we typically start with a large value of
$\alpha$ and make it smaller until the $\chi^2$ no longer varies.  To
test the effect of smoothing we run models such that the iterations
stop when $\alpha=0.01$, a reasonable value based on \citet{tho05}.

Using only our HET data we ran our no-halo and NFW-halo models with
and without smoothing using a coarser grid in parameter space.  We
find that smoothing does not alter the results.  All of the models
have a lower $\chi^2$ using only the HET data than the main results of
our paper which use both HET and SAURON data.  The models with
smoothing have a larger $\chi^2$ than without smoothing (see Table
\ref{tab:modressmooth}), but the $\Delta\chi^2$ between different halo
models remains the same.  The best-fitted NFW dark halo parameters are
consistent within the errors.  The internal moments are also
consistent with the unsmoothed models within the errors.  Using an
estimate by eye, the smoothed model's DF (plotted as $I_3$ versus
$L_z$ in $E$ bins) looks similar to the unsmoothed model's DF when
smoothed.

\begin{deluxetable}{cccccc}
\tablewidth{0pt}
\tablecaption{Smoothing Model Results\label{tab:modressmooth}}
\tablehead{
\colhead{halo} & \colhead{smoothing} & \colhead{$\chi^2$} & \colhead{$r_s$} & \colhead{$c$} & \colhead{$\rho$}\\
 & & & $kpc$ & & $M_{\odot}/pc^3$\\
(1) & (2) & (3) & (4) & (5) & (6)}
 \startdata
none & no & 940.857  & \nodata  & \nodata  & \nodata \\
none & yes & 1033.73  & \nodata  & \nodata  & \nodata \\
NFW & no & 766.47 & 1050 & 2.32 & $1.14\times10^{-4}$ \\
NFW & yes & 852.77 & 800 & 2.70 & $1.56\times10^{-4}$ \\
 \enddata
\tablecomments{(1) Dark halo density profile. (2) Smoothing or no smoothing. (3) $\chi^2$ of best-fitted model.  These $\chi^2$ values are lower than those of Table \ref{tab:modres} because the models fit fewer data points (HET data only) than the models in Table \ref{tab:modres} (HET and SAURON). (4) Scale radius of best-fitted model. (5) NFW concentration parameter determined from scale density. (6) NFW Scale density $\rho_{crit}\delta_c$.}
 \end{deluxetable}
 
We therefore determine that adding smoothing via maximum entropy does
not alter the measured halo, internal moments, or overall DF shape.
We also note that these results using only our HET data are consistent
with those using both HET and SAURON presented throughout this paper.

\subsection{Planetary Nebula Data}\label{pne}

\begin{figure}
\plotone{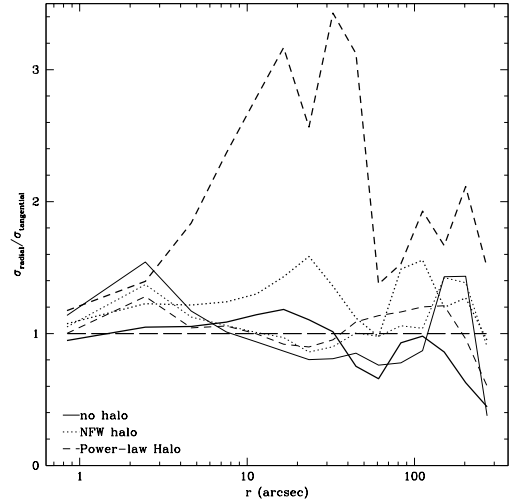}
\caption{Ratio of radial to tangential dispersion (an average of $\sigma_{\theta}$ and $\sigma_{\phi}$, including streaming motion) along the major axis for models with only planetary nebula data (thick lines) and the best-fitted no halo, NFW halo, and power-law halo derived from the stellar kinematics.  All three models are consistent with the PN data, in terms of $\chi^2$, with the dark halo models providing a slightly better fit.  Thin lines show results for stellar data as comparison.
\label{fig:intmompne}}
\end{figure}

We model the NGC~821 planetary nebula data of \citet{rom03} with the
best-fitted halos from the stellar data.  We are not trying to
constrain models using this data, but rather are interested in what
orbital properties the planetary nebulae would require given the
potential derived from the stellar data.  In doing this we assume that
the potential derived from stars is correct and that the planetary
nebulae are distributed in the same way as the stars.  This assumption
may not be realistic, as \citet{dek05} predicts that it is the
densities, not the anisotropies, that differ.  Figure
\ref{fig:intmompne} shows the ratio of radial to tangential dispersion
for the models with no dark halo and best-fitted NFW and power-law
halos.  As expected from the results of \citet{rom03}, the model with
no dark halo is roughly isotropic throughout, and tangential at large
radii.  The best-fitted NFW model requires radial orbits throughout
and the best-fitted power-law halo requires extremely radial orbits, with $\sigma_{radial}/\sigma_{tangential}$ of over $3$ (corresponding to a $\beta$ of 0.9).
This again demonstrates the strong mass-anisotropy degeneracy in
dynamical studies.  All three models are an excellent fit to the data,
although there is a preference for a dark halo, but it is not
statistically significant.

\section{Conclusions}\label{disc}

We present kinematics of NGC~821 to over 2 effective radii using
long-slit spectroscopy from the Hobby-Eberly Telescope and find that
our measured stellar line-of-sight velocity distributions are larger
than the planetary nebulae measurements of \citet{rom03} at large
radii.

Regardless of the density profile used, we are able to constrain the
enclosed mass of NGC~821 within our kinematic data ($\sim 2 R_e$) as
roughly $2\times10^{11}M_{\sun}$, equally divided between stars and
dark matter.  At $1R_e$ the ratio of dark matter halo mass to total
mass is $0.19$ for the best-fitted NFW halo profile and $0.13$ for the
best-fitted power-law halo profile.

We find that the best-fitted model of the dark halo in NGC~821 has a
nearly flat power-law density profile.  This dark halo gives a better
fit than both the NFW halo models and models without a dark halo at a
greater than 99\% confidence level.  This slope is somewhat
unexpected, and is strongly inconsistent with halo profiles with inner
slopes greater than one \citep[e.g. isothermal,][]{her90,moo99}, and
may lend support to halos with a flat inner slope \citep[e.g. cored
isothermal, logarithmic potential, and][]{bur95}.  Additionally, one
would expect that adiabatic contraction would create even steeper
inner halo profiles \citep{blu86,gne04}, which is in conflict with our
result.  This halo result is driven by the data at large radii.

Our NFW $\chi^2$ space shows a degeneracy in radius and density as
expected.  This degeneracy is slightly tilted from the expected NFW
correlations.  Constraining these NFW radius and density parameters
using a single concentration parameter could lead to biased results.

In addition to having a significantly poorer fit, the models without a
dark halo show tangential anisotropy at large radii.  This may be an
indication that a dark halo is necessary because the radial component
of the velocity dispersion may need to be artificially decreased at
large radii in order to create a smaller total velocity dispersion
that can be reproduced by a haloless model.  The best-fitted dark halo
model shows a radial bias in the $\theta$ direction at all radii.
However we do show that the velocities in the $\phi$ direction are
greater than the radial component.  If the planetary nebulae are on
radial orbits, that would explain why our measured stellar velocity
dispersions are larger than the reported planetary nebulae dispersions
\citep{rom03}.  We show this by modeling the planetary nebula data
assuming the potential of our best-fitted halo models from the stellar
kinematics.  We find that the planetary nebulae do require radially
anisotropic orbits to match the best-fitted halo potentials.

\acknowledgments

The authors would like to thank Tim de Zeeuw for his many helpful
comments and discussions regarding the manuscript. We are very
grateful to David Fisher for providing the surface brightness
profile. K.G. acknowledges the support of the Texas Advanced Research
Program under grant 003658--0243-2001 and NSF-CAREER grant
AST03-49095.  The authors acknowledge use of the computational
resources at The University of Texas at Austin's Texas Advanced
Computing Center (\url{http://www.tacc.utexas.edu}) for the research
reported in this paper.  The Hobby-Eberly Telescope (HET) is a joint
project of the University of Texas at Austin, the Pennsylvania State
University, Stanford University, Ludwig-Maximilians-Universit\"at
M\"unchen, and Georg-August-Universit\"at G\"ottingen. The HET is
named in honor of its principal benefactors, William P. Hobby and
Robert E. Eberly.

\bibliographystyle{apj}
\bibliography{references}  

\clearpage

\clearpage

\clearpage

\clearpage

\clearpage

\end{document}